\documentclass[twoside]{dis07}
\usepackage[latin1]{inputenc}
\usepackage[dvips]{graphicx,epsfig,color}
\usepackage{wrapfig,rotating}
\usepackage{amssymb,amsmath,array}
\usepackage{pst-node,lscape,amsfonts,amsbsy,graphics,lscape,mathrsfs,dsfont}
\usepackage[T1]{fontenc}
\usepackage[latin1]{inputenc}
\usepackage{subfigure}

\newcommand{\Tr}{\mathrm{Tr}}

\newcommand{\bea}{\begin{eqnarray}}
\newcommand{\eea}{\end{eqnarray}}
\newcommand{\bfl}{\begin{flushleft}}
\newcommand{\efl}{\end{flushleft}}
\newcommand{\bm}[1]{\mbox{\boldmath $#1$}}
\begin{document}
\title{Flavor Dependence of T-odd PDFs}

\author{Leonard P. Gamberg$^1$, Gary R. Goldstein$^2$, and Marc Schlegel$^3$
%
%
\vspace{.3cm}\\
%
1- Penn State University-Berks -Department of Physics and Astronomy \\
Reading, PA 19610 - USA 
%
\vspace{.1cm}\\
2- Tufts University - Department of Physics and Astronomy \\
Medford, MA 02155 - USA\\
3- Thomas Jefferson National Accelerator Facility - Theory Division\\
Newport News, VA 23608- USA\\
}

\maketitle

\begin{abstract}
The flavor dependence of the
 naive time reversal odd (``T-odd'') parton distributions 
 for  $u$- and $d$-quarks are explored in the spectator model. 
The flavor dependence of $h_{1}^{\perp}$ is of significance for the analysis 
of the azimuthal $\cos(2\phi)$
asymmetries in unpolarized SIDIS and DY-processes, as well as for
the overall physical understanding of the distribution of transversely
polarized quarks in unpolarized nucleons. 
As a by-product of the formalism,
 we calculate the chiral-odd but ``T-even''
function $h_{1L}^{\perp}$ which enables
us to present a prediction for the single spin asymmetry $A_{UL}^{\sin(2\phi)}$
for a longitudinally polarized target in SIDIS. 
\end{abstract}

Naive time reversal-odd ({}``T-odd'') transverse momentum dependent
(TMD) parton distributions (PDFs) have gained considerable attention
in recent years. Theoretically 
 they  can account for  non-trivial transverse 
spin  and momentum correlations 
such as single spin asymmetries (SSA) in hard scattering processes
when transverse momentum scales  are on the order of that 
of quarks in hadrons, namely 
$P_T\sim k_\perp  \ll \sqrt{Q^2}$.  A prominent example is the Sivers
function $f_{1T}^{\perp}$ \cite{Sivers:1989cc} which
explains the observed SSA in semi-inclusive deep inelastic 
scattering (SIDIS) for a
transversely polarized proton target by the HERMES 
collaboration \cite{Airapetian:2004tw}. 
It describes correlations of the intrinsic quark transverse momentum and the transverse nucleon spin.
The corresponding SSA on a deuteron target measured
by COMPASS \cite{Alexakhin:2005iw} vanishes, indicating 
a flavor dependence of the Sivers function. 
Another  leading twist ``T-odd'' parton distribution,
the chiral-odd Boer-Mulders function
$h_{1}^{\perp}$~\cite{Boer:1997nt} correlates the transverse spin of a quark
with its transverse momentum within the nucleon. 
We focus on the flavor dependence of these 
{}``T-odd'' functions where for example $h_{1}^{\perp}$
is  important for the analysis
of the azimuthal $\cos(2\phi)$ asymmetry in unpolarized SIDIS and
Drell-Yan~\cite{Goldstein:2002vv,Gamberg:2003ey}.
We also consider the flavor dependence of the ``T-even'' function 
$h_{1L}^\perp$,  which is of interest in the transverse momentum and quark spin correlations
in a longitudinally polarized target~\cite{Kotzinian:1997wt}.

Considerable understanding of TMDs and fragmentation functions (FF)
have  been gained from model calculations using the spectator framework
\cite{Brodsky:2002cx,Ji:2002aa,Goldstein:2002vv,Gamberg:2003ey,Bacchetta:2003rz,Gamberg:2006ru}.

\begin{figure}
\begin{center}\subfigure[Nucleon-Diquark-Quark
 vertex]{\includegraphics[scale=0.2]{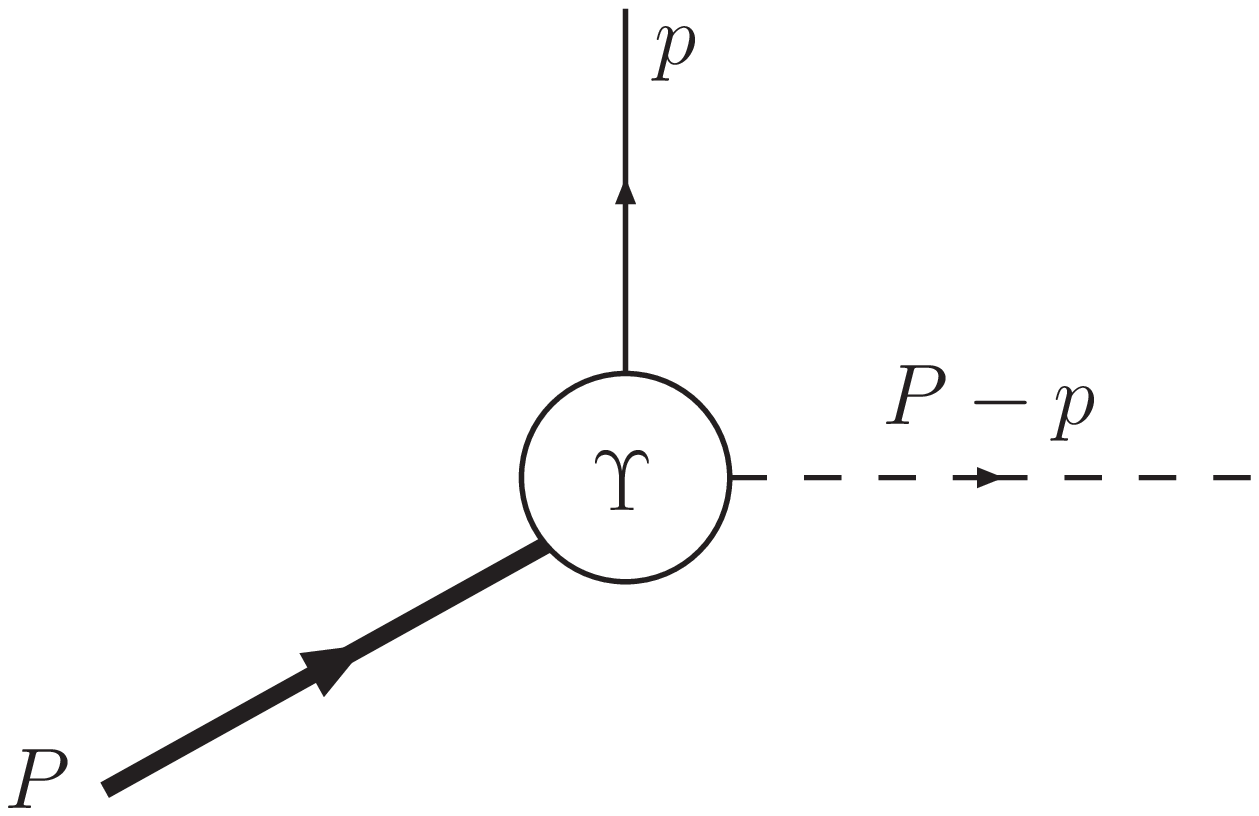}}
~~~~~~~~~~~~\subfigure[Diquark-gluon vertex]
{\includegraphics[scale=0.2]{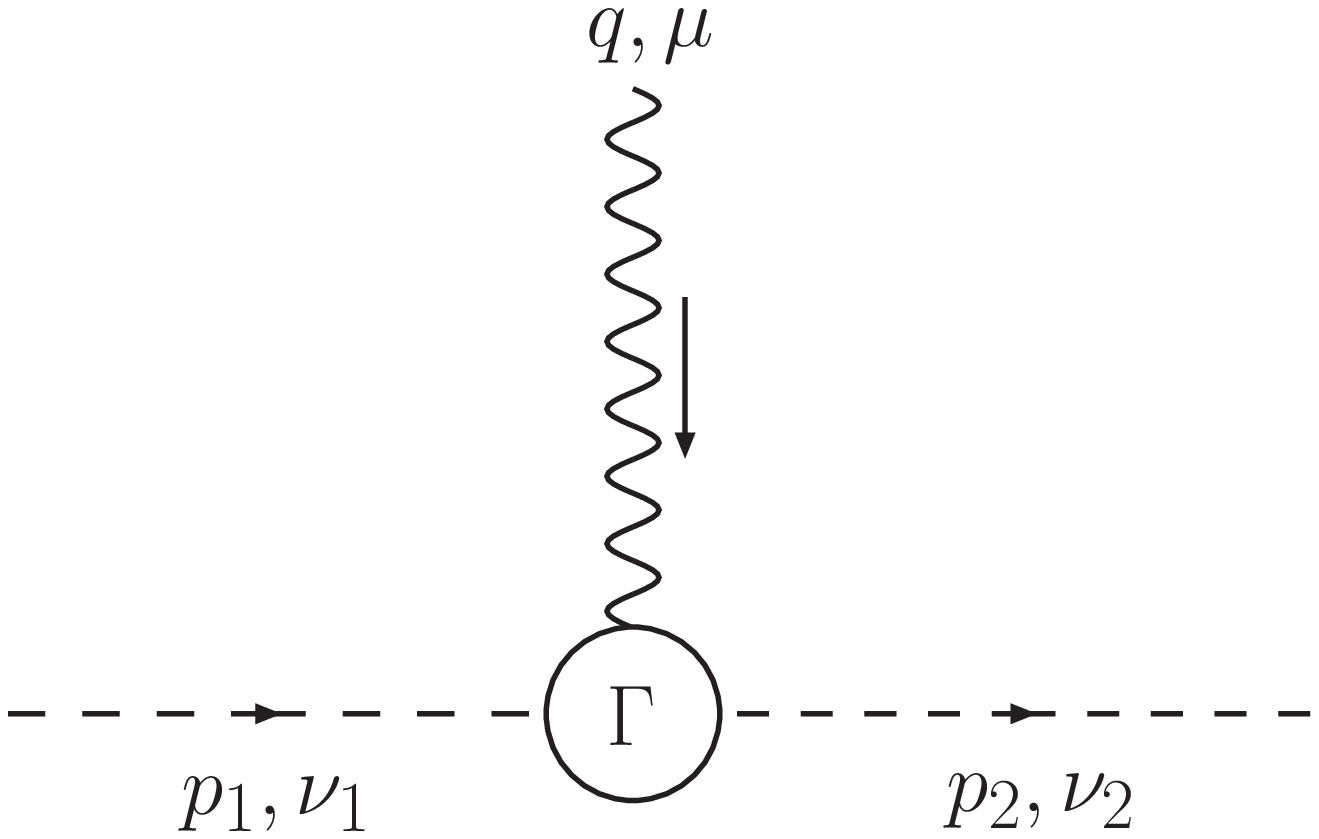}}
~~~~~~~\subfigure[Box-graph hermitian conjugated]{\includegraphics[scale=0.2]{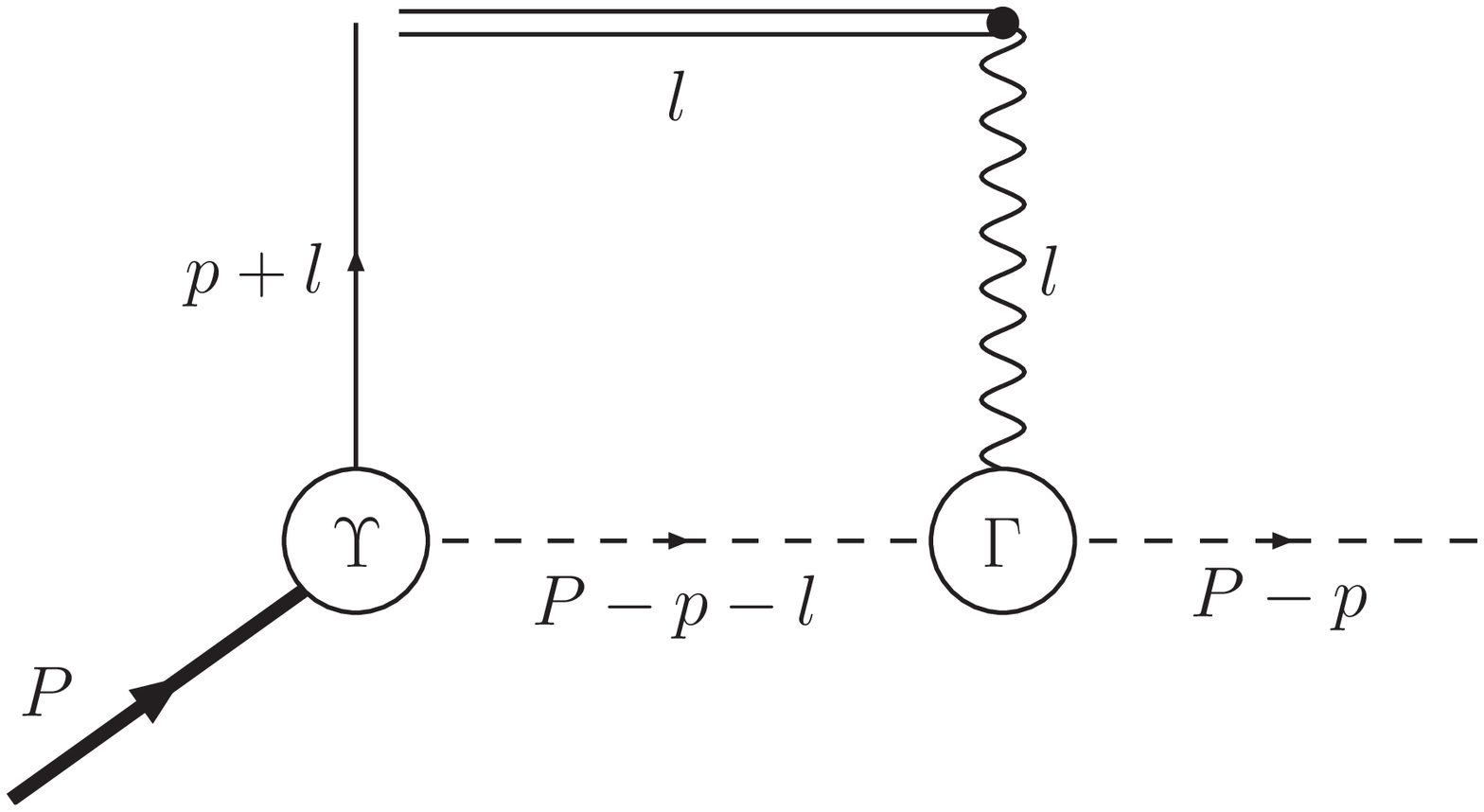}}
\end{center}
\vspace{-0.4cm}
\caption{\small{Different vertices for the axial-vector diquark and
contribution of the gauge link in the one-gluon approximation.} 
\vspace{-0.4cm}
\label{cap:Different-vertices-for}}
\end{figure}

In this formalism we
start (cf. \cite{Jakob:1997wg}) from the definition of the 
unintegrated
color gauge invariant quark-quark correlator 
which contains  the gauge link  indicated by the Wilson line,
$\mathcal{W}[a\,|\, b]$,
and work in Feynman gauge in which the {\underline{transverse}} Wilson line
vanishes \cite{Belitsky:2002sm}.
In the diquark model the sum over the  complete set of intermediate on-shell
states in the definition of the correlator is represented 
by a single one-particle diquark state $|\, dq;\, p_{dq},\lambda\rangle$, where $p_{dq}$ is the diquark
momentum and $\lambda$ its polarization. The diquark is {}``built''
from two valence quarks which can  be scalar-spin 0 
or axial vector-spin 1. The unintegrated correlator is then given by
\begin{eqnarray}
\Phi_{ij}(p;P,S) \hspace{-0.2cm}& = & \hspace{-0.2cm}\sum_{\lambda}\frac{\delta((P-p)^{2}-m_{s}^{2})\Theta(P^{0}-p^{0})}{(2\pi)^{3}}\,\langle P,S|\,\bar{\psi}_{j}(0)\,\mathcal{W}[0\,|\,\infty,0,\vec{0}_{T}]\,|\, dq;\, P-p,\lambda\rangle\times\nonumber \\
 &  & \langle\, dq;\, P-p,\lambda|\,\mathcal{W}[\infty,0,\vec{0}_{T}\,|\,0]\,\psi_{i}(0)\,|P,S\rangle.\label{eq:Correlator}\end{eqnarray}
 The essence of the diquark spectator model is to calculate the matrix
elements in Eq. (\ref{eq:Correlator}) by the introduction of effective
The nucleon-diquark-quark vertices
$\Upsilon_{s}(N)$ and $\Upsilon_{ax}^{\mu}(N)$
are represented in  Fig. \ref{cap:Different-vertices-for} (a).  
For example, the matrix element 
for the axial vector diquark is
 \begin{eqnarray}
\langle adq;\, P-p;\lambda|\,\psi_{i}(0)\,|P,S\rangle 
= i\frac{g_{ax}(p^{2})}{\sqrt{3}}\varepsilon_{\mu}^{*}(P-p;\lambda)\frac{\left[(\slash\hspace{-.2cm} p+m_{q})\gamma_{5}\left[\gamma^{\mu}-R_{g}\frac{P^{\mu}}{M}\right]u(P,S)\right]_{i}}{p^{2}-m_{q}^{2}+i0},\label{eq:f1MEax}\end{eqnarray}
 where the polarization vector of the axial-vector diquark is 
$\varepsilon_{\mu}$, $u(P,S)$ denotes the nucleon spinor and
$M$ and $m_{q}$ are nucleon and quark masses, respectively. 
 The unpolarized TMD $f_{1}$ is obtained by  inserting these expressions
into Eq. (\ref{eq:Correlator})
and projecting from the quark-quark correlator 
\begin{equation}
f_{1}(x,\vec{p}_T^2)=\frac{1}{4}\int dp^{-}\,\left(\Tr\left[\gamma^{+}\Phi(p;P,S)\right]+\Tr\left[\gamma^{+}\Phi(p;P,-S)\right]\right)\bigg|_{p^{+}=xP^{+}},\end{equation}
We have also calculated
 the distribution of transversely
polarized quarks in a longitudinally polarized target, $h_{1L}^{\perp}$
by replacing $\Gamma^+$ with $\Gamma^+\Gamma^i\Gamma_5$ and 
$S$ by $S_L$,
the spin 4-vector in longitudinal direction. Their analytic expressions in the scalar and axial vector diquark sectors
are given in~\cite{Gamberg:2007wm}.

In the spectator framework the ``T-odd'' TMDs \cite{Brodsky:2002cx} are
generated by the gauge link \cite{Ji:2002aa,Goldstein:2002vv,Gamberg:2003ey}.
The leading contribution, 
arising from  the interference  between
tree- and box graph
which contains an imaginary part necessary for ``T-odds'',
is represented in Fig. \ref{cap:Different-vertices-for} (c) in which the double
line is an eikonal,  and  $l$ is the loop momentum. For an axial-vector diquark
we   model the composite 
nature of the diquark through an anomalous magnetic
moment $\kappa$~\cite{Goldstein:1979wb}. 
In the notation of Fig. \ref{cap:Different-vertices-for}
(b) the gluon-diquark axial diquark vertex is 
\begin{equation}
\Gamma_{ax}^{\mu\nu_{1}\nu_{2}}=-ie_{dq}\left[g^{\nu_{1}\nu_{2}}(p_{1}+p_{2})^{\mu}+(1+\kappa)\left(g^{\mu\nu_{2}}(p_{2}+q)^{\nu_{1}}+g^{\mu\nu_{1}}(p_{1}-q)^{\nu_{2}}\right)\right].\end{equation}
 For $\kappa=-2$ the vertex $\Gamma_{ax}$ reduces to the standard
$\gamma WW$-vertex. We express the matrix elements including the gauge link,
$\langle dq;\, P-p|\,\mathcal{W}[\infty,0,\vec{0}_{T}\,|\,0]\,\psi_{i}(0)\,|P,S\rangle $
 in the one gluon approximation~\cite{Gamberg:2007wm}. 
Projecting the Boer-Mulders function, 
\begin{equation}
\epsilon_{T}^{ij}p_{T}^{j}h_{1}^{\perp}(x,\vec{p}_{T}^{2})=\frac{M}{4}\int dp^{-}\left(\Tr\left[\Phi_{\mathrm{unpol}}(p,S)i\sigma^{i+}\gamma_{5}\right]+\Tr\left[\Phi_{\mathrm{unpol}}(p,-S)i\sigma^{i+}\gamma_{5}\right]\right), \label{bm}\end{equation}
where $\epsilon_{T}^{ij}\equiv\epsilon^{-+ij}$
the axial-vector diquark contribution is given by 
the expression,
\begin{eqnarray}
\epsilon_{T}^{ij}p_{T}^{j}h_{1}^{\perp,ax}(x,\vec{p}_{T}^{2}) & = &
 -\frac{e_{q}e_{dq}}{8(2\pi)^{3}}\frac{1}{\vec{p}_{T}^{2}+\tilde{m}^{2}}\frac{M}{P^{+}}\;
\int\frac{d^{4}l}{(2\pi)^{4}}\;\bigg\{\frac{1}{3}g_{ax}\left((l+p)^{2}\right)g_{ax}^{*}
\left(p^{2}\right)
\times\nonumber \\ &  & \hspace{-3cm} \mathcal{D}_{\rho\eta}(P-p-l)\left(\sum_{\lambda}\varepsilon_{\sigma}^{*}(P-p;\lambda)\varepsilon_{\mu}(P-p;\lambda)\right)\times\nonumber \\
 &  & \hspace{-1cm}\frac{\left[g^{\sigma\rho}\, v\cdot(2P-2p-l)+(1+\kappa)\left(v^{\sigma}\,(P-p+l)^{\rho}+v^{\rho}\,(P-p-2l)^{\sigma}\right)\right]}{\left[l\cdot v+i0\right]\left[l^{2}-\lambda^{2}+i0\right]\left[(l+p)^{2}-m_{q}^{2}+i0\right]}\times\nonumber \\
 &  & \hspace{-3.5cm}\Tr\Big[\left(\slash\hspace{-.2cm}
			    P+M\right)\left(\gamma^{\mu}-R_{g}
\frac{P^{\mu}}{M}\right)\left(\slash\hspace{-.2cm} p-m_{q}\right)\gamma^{+}\gamma^{i}
 \left(\slash\hspace{-.2cm} l+\slash\hspace{-.2cm} p+m_{q}\right)\left(\gamma^{\eta}+R_{g}\frac{P^{\eta}}{M}\right)\gamma_{5}\Big]\bigg\}+\;\mathrm{h.c.}\;.\label{eq:StartingFormula}\end{eqnarray}
$\mathcal{D}(P-p-l)$ denotes the propagator of the
 axial-vector diquark.  Since the numerator in Eq. (\ref{eq:StartingFormula}) contains at most
the loop momentum to the fourth power we can write it in the following
manner, $\sum_{i=1}^{4}N_{\alpha_{1}...\alpha_{i}}^{(i)}l^{\alpha_{1}}...l^{\alpha_{i}}+N^{(0)}.\label{eq:NumDecomp}$
 The (real) coefficients (tensors) $N_{\alpha_{1}...\alpha_{i}}^{(i)}$
depend only on external momenta and 
can be computed in a straight-forward manner. The integration over the light cone components,  $l^{+}$ and
$l^{-}$, are easily performed;  however, calculating 
the $l^{+}$-integral  results in an  integral that is potentially ill-defined.  This happens 
when $g(p^{2})$ is a holomorphic function in $p^{2}$
 and at least one of the Minkowski indices
is light-like in the minus direction, e.g. $\alpha_{1}=-$,
$\alpha_{2},...,\alpha_{i}\in\left\{ +,\perp\right\} $ resulting in an
integral of the form $\int dl^{+}\,\delta(l^{+})\Theta(-l^{+})$, 
implying that $l^{+}=0$ and $l^{-}=\infty$. This 
signals the existence of a light cone divergence in Ref.~\cite{Gamberg:2006ru}. 

\begin{figure}[top]
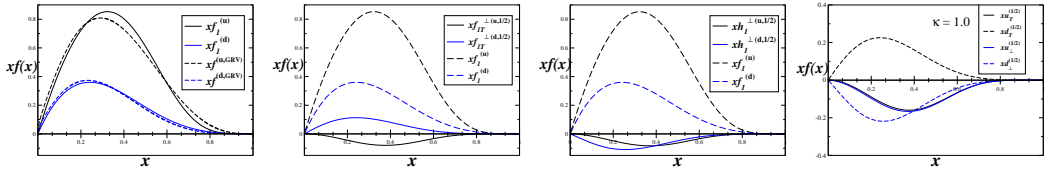

\begin{center}
\subfigure{\includegraphics[scale=0.1325]{un_axialpubgrv.eps}}
~\subfigure{\includegraphics[scale=0.1325]{siv_axialpub.eps}}
~\subfigure{\includegraphics[scale=0.1325]{bm_axialpub.eps}}
\subfigure{\includegraphics[scale=0.1325]{sivbm1_axialpub.eps}}
\vspace{-0.4cm}
\caption{\small{
The unpolarized up and down quark distributions functions (left) 
versus $x$ compared to
the low scale parameterization of GRV
~\cite{Gluck:1998xa}.  The half moment of the  Sivers function (center-left)
and Boer-Mulders function (center-right)
and the unpolarized up and down quark distributions ($\kappa=1.0$).
The half-moment (left) and first moments (right) of the Boer-Mulders  
and Sivers  functions versus $x$  
fitted to extractions from data were presented in Ref. \cite{Anselmino:2005an}
\vspace{-0.4cm}
}\label{cap:xfg}}
\end{center}
\end{figure}

One can handle the light cone divergences
by introducing phenomenological form factors with additional poles~\cite{Gamberg:2007wm},
\begin{equation}
g_{ax}(p^{2})=\frac{(p^{2}-m_{q}^{2})f(p^{2})}{\left[p^{2}-\Lambda^{2}+i0\right]^{n}}. \label{eq:FFn}\end{equation}
For $n\geq3$ there are  enough powers of $l^{+}$  to eliminate this divergence.
  $f(p^{2})$ is a covariant Gaussian~\cite{Gamberg:2007wm}
which cuts off the $p_{T}$ integrations and  $\Lambda$ is an arbitrary
mass scale fixed by fitting $f_{1}$ to
data. Similarly, the Sivers-function is projected
from the  trace of the quark-quark correlator (\ref{eq:Correlator})
(see e.g. \cite{Bacchetta:2006tn}),\begin{equation}
2S_{T}^{i}\epsilon_{T}^{ij}p_{T}^{j}f_{1T}^{\perp}(x,\vec{p}_{T}^{2})=\frac{M}{2}\int dp^{-}\,\left(\Tr\left[\gamma^{+}\Phi(p;P,S_{T})\right]-\Tr\left[\gamma^{+}\Phi(p;P,-S_{T})\right]\right)\bigg|_{p^{+}=xP^{+}}.
\label{siv}
\end{equation}  
It is well-known \cite{Goldstein:2002vv} that in the scalar diquark
approximation the $h_1^\perp$  and $f_{1T}^\perp$
coincide.  By contrast the different Dirac structure
for the chiral even $f_{1T}^\perp$
 and chiral odd  $h_1^\perp$ 
in the axial-vector diquark sector, Eq. (\ref{bm}) and 
(\ref{siv})
respectively, lead to different coefficients in the decomposition
$N_{\alpha_{1}...\alpha_{i}}^{(i)}$~\cite{Gamberg:2007wm}.
We  fix most of the model parameters such as masses and normalizations
by comparing the model result for the unpolarized  $f_{1}$
for $u$ and $d$ quarks 
to  the low-scale ($\mu^{2}=0.34{\textrm{GeV}}$) data parameterization
of GRV~\cite{Gluck:1998xa} see Fig. \ref{cap:xfg}. 
Note that PDFs
for $u$ and $d$ quarks are given by linear combinations of PDFs
for an axial vector and scalar diquark, $u=\frac{3}{2}f^{sc}+\frac{1}{2}f^{ax}$
and $d=f^{ax}$ \cite{Jakob:1997wg,Bacchetta:2003rz}. 
 For ``T-odd'' PDFs we fix the sign and the strength
of the final state interactions,
the product of the charges of the diquark and quark,
 by comparing 
  $f_{1T}^\perp$ for $u$ and $d$ quarks in the
diquark model  with the  existing data parameterizations (see Ref. \cite{Anselmino:2005an}).
The ``one-half'' moments $q_T^{(1/2)}$ 
(where $q={u,d}$) of the Sivers and Boer-Mulders functions
$$\frac{1}{2}q_T^{(1/2)}(x)=f_{1T}^{\perp(q,1/2)}(x)=\int{d^2 p_T\,\frac{|\vec{p}_T|}{2M}\,f_{1T}^{\perp (q)}(x,\vec{p}_T^2)},
$$ are displayed 
along with the unpolarized $u$ and $d$ quark distributions in Fig. \ref{cap:xfg}. 
The {}``one-half'' and first moments~\cite{Gamberg:2007wm} of the up and
down quark Sivers functions are negative and positive respectively
while the up and down quark Boer-Mulders functions are {\em both negative}
over the full range in Bjorken-$x$. 
We also note that the  $u$-quark Sivers function and 
Boer-Mulders function are nearly equal, even with the inclusion of the 
axial vector spectator diquark. 

\begin{figure}
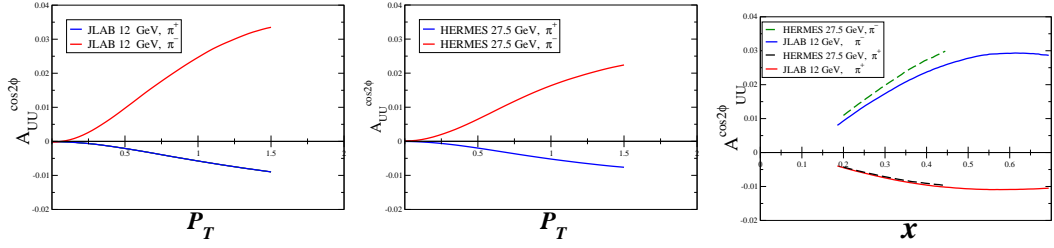

\begin{center}
~\subfigure{\includegraphics[scale=0.18]{cospt_pub.eps}}
~\subfigure{\includegraphics[scale=0.18]{cospt_pubh.eps}}
~\subfigure{\includegraphics[scale=0.18]{cosx_pub.eps}} 
\end{center}
\vspace{-0.4cm}
\caption{\small{The $\cos2\phi$ asymmetry for 
and $\pi^{\pm}$ as a function of $P_{T}$ at JLAB $12\textrm{GeV}$ and HERMES (center) kinematics.
Right: $\cos2\phi$ asymmetry for $\pi^{\pm}$ as a function of $x$ at JLAB $12\textrm{GeV}$ 
and HERMES kinematics. 
} 
\vspace{-0.4cm}
\label{asym:pt}}
\end{figure}

\begin{figure}
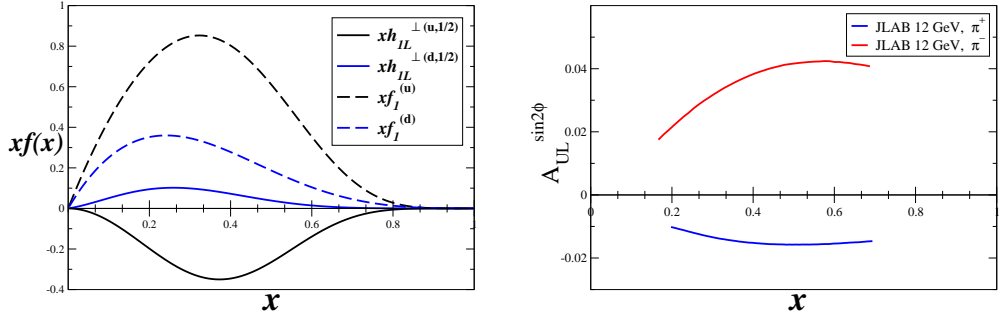

\begin{center}
\subfigure{\includegraphics[scale=0.25]{h1Lnew.eps}}
~~~~~\subfigure{\includegraphics[scale=0.25]{sinx_pub.eps}}
\end{center}
\vspace{-0.4cm}
\caption{\small{Left Panel:
The half-moment of $xh_{1L}^{\perp(1/2)}$
 versus $x$ compared to the unpolarized up and down quark
distribution functions. 
Right Panel: The $\sin2\phi$ asymmetry for $\pi^{+}$ and
$\pi^{-}$ as a function of $x$ at JLAB $12\textrm{GeV}$ 
kinematics. 
} \vspace{-0.4cm}
\label{h1L}}
\end{figure}

Having explored the flavor dependence of the $h_{1}^{\perp}$ we are
now in a position to extend early phenomenological work on ``T-odd'' contributions
to azimuthal asymmetries in SIDIS\cite{Gamberg:2003ey}. We
consider the spin independent double ``T-odd'' $\cos2\phi$ asymmetry
for $\pi^{+}$ and $\pi^{-}$ production.
We focus on the important contributions to the  cross section for unpolarized SIDIS 
\cite{Bacchetta:2006tn}\begin{eqnarray}
\frac{d\sigma}{dx\, dy\,\, dz\, d\phi_{h}\, dP_{h\perp}^{2}} \approx  
\frac{2\pi\alpha^{2}}{xyQ^{2}}\,\Big[\left(1-y+\frac{1}{2}y^{2}\right)\, F_{UU,T}\, 
+\,\left(1-y\right)\cos(2\phi_{h})\, F_{UU}^{\cos2\phi_{h}}\Big],\nonumber \end{eqnarray} 
where the structure function $F_{UU}^{\cos2\phi_{h}}$ 
involves  a convolution
of the Boer-Mulders and Collins fragmentation function 
 \bea
F_{UU}^{\cos 2\phi_h} = \mathcal{C}\biggl[
   - \frac{2  \hat{\bm{h}}\cdot \bm{k}_T\hat{\bm{h}}\cdot \bm{p}_T 
    -\bm{k}_T \cdot \bm{p}_T}{M M_h}
    h_{1}^{\perp } H_{1}^{\perp }\biggr],\label{FUUcos2phi} \eea
$\mathcal{C}$ is the convolution integral.  Our input for the Collins functions is based on recent work
in~\cite{Bacchetta:2007xx} where the Collins function was calculated in the spectator framework. It was assumed 
that $H_1^{\perp(dis-fav)}\approx -H_1^{\perp(fav)}$ in the pion sector, thereby satisfying the Schaefer-Teryaev sum rule \cite{Schafer:1999kn} locally.  
We estimate the azimuthal asymmetry $A_{UU}^{\cos2\phi}$ (cf. Eq. (\ref{FUUcos2phi})), where
$A_{UU}^{\cos 2\phi}\equiv \int  \cos 2\phi\ d\sigma /\int \ d\sigma$.  
In Fig.~\ref{asym:pt} we display the $A_{UU}^{\cos2\phi}(P_{T})$
in the range of HERMES~\cite{Airapetian:2004tw} and future JLAB kinematics~\cite{Clas:2006ka} 
 as well as  $x$ dependence in the 
range $0.5<P_{T}<1.5\,\, \rm{GeV/c}$. 
Having calculated the chiral-odd but ``T-even'' parton distribution
$h_{1L}^{\perp}$  we use this result together with the result of
Ref. \cite{Bacchetta:2007xx} for the Collins function to give a prediction
for the $\sin(2\phi)$  single spin asymmetry $A_{UL}$
for a longitudinally polarized target.  A decomposition into structure functions of the cross section of semi-inclusive
DIS for a longitudinally polarized target reads (see e.g. \cite{Bacchetta:2006tn})
\begin{eqnarray}
\frac{d\sigma_{UL}}{dx\, dy\, dz\, d\phi_{h}\, dP_{h\perp}^{2}} & \approx & \frac{2\pi\alpha^{2}}{xyQ^{2}}\, S_{\parallel}\,\Big[(1-y)\sin(2\phi_{h})\, F_{UL}^{\sin(2\phi)}+(2-y)\sqrt{1-y}\sin(\phi_{h})\, F_{UL}^{\sin\phi}\Big],\nonumber \\
\end{eqnarray}
$S_{\parallel}$ is the projection of the spin vector on the
direction of the virtual photon. In a partonic picture the structure
function $F_{UL}^{\sin(2\phi)}$ is a leading twist object (while
$F_{UL}^{\sin\phi}$ is sub-leading) and given by a convolution of
the TMD $h_{1L}^{\perp}$ and the Collins function (cf. \cite{Bacchetta:2006tn})
\begin{equation}
F_{UL}^{\sin(2\phi)}=\mathcal{C}\Bigg[-\frac{2\hat{\bm{h}}\cdot\bm{k}_{T}\,\hat{\bm{h}}\cdot\bm{p}_{T}-\bm{k}_{T}\cdot\bm{p}_{T}}{MM_{h}}\, h_{1L}^{\perp}H_{1}^{\perp}\Bigg]. \label{FULsin2phi}\end{equation} 
We display the results for the single spin asymmetry $A_{UL}^{\sin(2\phi)}$
in Fig.~\ref{h1L} using the kinematics of the upcoming JLab 12 GeV upgrade. 
We note that the $\pi^-$ asymmetry is large and positive due to the model assumption $H_1^{\perp(dis-fav)}\approx -H_1^{\perp(fav)}$. 
This asymmetry has been measured at HERMES for longitudinally polarized
protons~\cite{Airapetian:1999tv} and deuterons~\cite{Airapetian:2002mf}.
The data show that for the proton target at HERMES 27.5 GeV kinematics both $\pi^+$ and $\pi^-$ 
asymmetries are consistent with 0 down to a sensitivity of about 0.01. 
These asymmetries could be non-zero, but with magnitudes less than 0.01 
or 0.02. These results are considerably smaller than our predictions for the JLab upgrade. 
For the deuteron target the results are  consistent with 0 for $\pi^+$ and $\pi^-$.  This SIDIS data for polarized deuterons could reflect the near cancellation 
of $u$- and $d$-quark $h_{1L}^{\perp}$ functions and/or the large unfavored Collins function contributions. 
There is also CLAS preliminary data~\cite{Avakian:2005ps} at 5.7 GeV that shows 
slightly negative asymmetries for $\pi^+$ and $\pi^-$ and leads to the extraction 
of a negative $xh_{1L}^{\perp(u)}$. This suggests that the unfavored Collins 
function (for $d\rightarrow\pi^+$) is not contributing much. 
Data from the upgrade should help resolve these phenomenological questions.

We have performed calculations of  transverse momentum
dependent parton distributions, including the Boer-Mulders function 
$h_{1}^{\perp}$, the Sivers function $f_{1T}^{\perp}$ 
along with $h_{1L}^{\perp}$ in the framework of an axial-vector and a scalar
diquark spectator model. The calculation of these functions in both
sectors allowed us to explore their flavor dependence, i.e. to compute 
$h_{1}^{\perp}$, $f_{1T}^{\perp}$ and
$h_{1L}^{\perp}$ for a $u$-quark and a $d$-quark.  We used these results
 along with  the Collins fragmentation function $H_{1}^{\perp}$.
to  estimate the azimuthal asymmetries  $A_{UU}^{(\cos(2\phi))}$ and $A_{UL}^{(\sin(2\phi))}$ in SIDIS.  In summary, a refined diquark spectator model, including axial vector di-quarks leads to 
both $u$- and $d$-quark ``T-odd'' TMDs and provides the ingredients for predicting a 
range of asymmetries for future experiments.  
\vspace{-0.08cm}
\bfl
{\bf Acknowledgments}
\efl
\vspace{-0.08cm}
This work
is supported in part by the U.S. Department of Energy under contracts,
DE-FG02-07ER41460 (LG), and DE-FG02-92ER40702 (GRG).
 

\begin{footnotesize}


\bibliographystyle{h-physrev3}
\bibliography{Referenzen}
%

\end{footnotesize}


\end{document}